\documentstyle[11pt,emulateapj,psfig]{article}

\def\etal{{\it et al.~}}
\def\lsim{\hbox{ \rlap{\raise 0.425ex\hbox{$<$}}\lower 0.65ex\hbox{$\sim$} }}
\def\gsim{\hbox{ \rlap{\raise 0.425ex\hbox{$>$}}\lower 0.65ex\hbox{$\sim$} }}

\def\f(h{\hbox{$~\!\!^{\rm h}$}}

\def\ale{\mathrel{\hbox{\rlap{\hbox{\lower4pt\hbox{$\sim$}}}\hbox{$<$}}}}
\def\age{\mathrel{\hbox{\rlap{\hbox{\lower4pt\hbox{$\sim$}}}\hbox{$>$}}}}

\lefthead{Djorgovski et al.} 
\righthead{Quasar Pair at z $\approx$ 5}

\begin{document}

\title{Discovery of a Clustered Quasar Pair at z $\approx$ 5: \\
       Biased Peaks in Early Structure Formation}
\footnote{
Based on observations obtained at the W.~M.~Keck Observatory
which is operated by the California Association for Research in Astronomy, a
scientific partnership among California Institute of Technology, the University
of California and the National Aeronautics and Space Administration.
\\
\null\qquad $^*$ Present address:  Dept. of Astronomy, Univ. of Illinois, Urbana, IL 61801.
}

\author{S. G. Djorgovski}

\affil{Palomar Observatory 105--24, California Institute of Technology,
       Pasadena, CA 91125, USA; {\tt george@astro.caltech.edu}}

\author{D. Stern}

\affil{Jet Propulsion Laboratory, 
       Pasadena, CA 91109, USA; {\tt stern@zwolfkinder.jpl.nasa.gov}}

\author{A. A. Mahabal, R. Brunner*}

\affil{Palomar Observatory 105--24, California Institute of Technology,
       Pasadena, CA 91125, USA; {\tt [aam,rb]@astro.caltech.edu}}


\begin{abstract}
We report a discovery of a quasar at $z = 4.96 \pm 0.03$ within a few Mpc
of the quasar SDSS 0338+0021 at $z = 5.02 \pm 0.02$.  The newly found quasar
has the SDSS $i$ and $z$ magnitudes of $\approx 21.2$, and an estimated
absolute magnitude $M_B \approx -25.2$.  The projected separation on the sky 
is 196 arcsec, and the redshift difference $\Delta z = 0.063 \pm 0.008$. 
The probability of finding this quasar pair by chance in the absence of
clustering in this particular volume is $\sim 10^{-4} - 10^{-3}$.  We conclude
that the two objects probably mark a large-scale structure, possibly a
protocluster, at  $z \approx 5$.  This is the most distant such structure
currently known.  
Our search in the field of 13 other QSOs at $z \gsim 4.8$ so far has not
resulted in any detections of comparable luminous QSO pairs, and it is thus
not yet clear how representative is this structure at $z \approx 5$.
However, along with the other evidence for clustering of quasars and young
galaxies at somewhat lower redshifts, the observations are at least
qualitatively consistent with a strong biasing of the first luminous
and massive objects, in agreement with general predictions of theoretical
models.  More extensive searches for clustered quasars and luminous galaxies 
at these redshifts will provide valuable empirical constraints for our
understanding of early galaxy and structure formation. 
\end{abstract}


\section{Introduction}

Quasars at large redshifts represent a powerful probe of structure formation
in the early universe.  
Dissipative mergers and tidal interactions during the early stages of galaxy
assembly may be fueling both bursts of star formation and early AGN activity
(see, e.g., \cite{sr98}, \cite{fhmm99}, \cite{msd00}, \cite{kh00},
\cite{gran01}, etc.).  This fundamental connection is supported by the
remarkable correlations between the masses of central black holes in
nearby galaxies, and the velocity dispersions and luminosities ($\sim$ masses)
of their old, metal-rich stellar populations (\cite{mag98}, \cite{fm00},
\cite{geb00}), and by the high metallicities observed in the high-$z$ quasar
spectra (\cite{hf99}, and references therein). 
Quasars can also be used directly to probe evolution of large-scale structure
out to high redshifts, as demonstrated clearly by the 2QZ survey 
(\cite{croom01}, \cite{hoy02}).

There should also be a fundamental connection between the formation of galaxies
and the large-scale density field in the early universe.  The highest density
peaks, where presumably the first luminous objects formed, should be
strongly clustered, due to biasing (\cite{kai84}).  This is a generic and
robust prediction for essentially every model of structure formation,
independent of any astrophysical details of galaxy formation. 

Luminous high-$z$ quasars are likely situated in massive hosts (\cite{tur91}).  
Such massive halos should be rare, and may be associated with $\sim 4$ to
$5$-$\sigma$ peaks of the primordial density field (\cite{er98}, \cite{ck89},
\cite{ns93}).  High-$z$ quasars can thus be used as biased tracers of the early
large-scale structure, possibly marking the cores of future rich clusters. 

A search for protoclusters around known high-$z$ objects such as quasars thus
provides an important test of our basic ideas about the biased galaxy formation.
We have conducted a search for clustered protogalaxies and AGN in the fields of
selected quasars at $z > 4$.  Preliminary results from our program have been
described by \cite{djorg99} and \cite{dj+99}, and a complete account will be
presented elsewhere. 

Here we present the discovery of a clustered quasar pair at 
$z \approx 5$, which we interpret as a signature of a primordial large-scale
structure, possibly a core of a forming rich cluster.  This is the most distant
large-scale structure currently known.  One of the quasars, SDSS 0338+0021, was
discovered by \cite{fan99}; the other was found in our much deeper search in
its vicinity.

\section{Observations and Data Reductions}

Deep images of the field were obtained at the Palomar 200-inch Hale telescope
using the prime-focus $Cosmic$ imager (\cite{kel98}).  The instrument FOV is
9.7 arcmin square, with 0.286 arcsec pixels.  Multiple dithered exposures
totalling 1800 s, 3000 s, and 7200 s were obtained in Gunn $r$, $i$, and $z$
filters respectively on 29 and 30 November 2000 UT, and of 3600 s, 1800 s, and
2400 s in Gunn $g$, $r$, and $z$ filters respectively on 31 December 2000 UT,
all in good conditions.  The data were reduced using standard procedures.
Limiting magnitudes (3-$\sigma$ in a 3 arcsec apertures) are
$r_{lim} \approx 25.9$, $i_{lim} \approx 25.5$, and $z_{lim} \approx 23.1$ mag
in the SDSS ($\sim AB_\nu$) photometric system.
Multi-color selection was used to identify candidates for objects at 
$z \gsim 5$, as illustrated in Fig. 1.  Details of the data reduction and
candidate selection procedure will be presented elsewhere.

The first set of imaging data was used to select candidates for multi-slit
spectroscopy, and spectra of several objects, including the known QSO (SDSS
0338+0021), were obtained at the WMKO Keck-I 10-m telescope on 29 December 2000
UT in good conditions.  We used the LRIS instrument (\cite{occ+95}) with a 400
lines mm$^{-1}$ grating ($\lambda_{\rm blaze} = 8500$\AA) and a GG495 order
sorting filter, through 1.2 arcsec wide slitlets, with a mean dispersion in 
the wavelength region of interest of $\approx 1.86$ \AA\ pixel$^{-1}$.  
Two exposures of 1800 s each were obtained through a single slitmask at a 
PA = 340$^\circ$, close to the parallactic angle at the time.  Exposures of
arc lamps were used for wavelength calibration, with the flexure shifts
corrected using the measurements of selected night sky lines.  An average
of archival response curves for this grating was used for the flux calibration.

Only one high-priority color-selected candidate, which we designated RD 657
(for ``red dropout'' and a serial number in our CCD object catalog), could
be accommodated on this slitmask.  It turned out to be a quasar at $z \approx
5$.  The position of this object (J2000) is: 
$$\alpha ~= ~03^h ~38^m ~30.03^s  ~~~~ 
\delta ~= ~+00^\circ ~18^{\prime} ~40.4^{\prime\prime},$$
giving the projected separation from SDSS 0338+0021 of 196 arcsec.
Setting the photometric zero-points on the magnitudes of SDSS 0338+0021
published by \cite{fan99}, i.e., 
$r = 21.70$,
$i = 19.96$, and
$z = 19.74$,
the magnitudes of the 
new QSO in the same SDSS system are
$r = 23.01$,
$i = 21.24$, and
$z = 21.16$ mag, with estimated uncertainties of $\sim 0.1$ mag,
making it about 3.5 times fainter than the SDSS QSO.
Fig. 2 shows the finding chart for the field, with both quasars marked.

Additional long-slit spectra of both objects were obtained using LRIS on 31
December 2000 UT in good conditions.  We used a 600 lines mm$^{-1}$ grating
($\lambda_{\rm blaze} = 7500$\AA) through a 1.0 arcsec wide slit, covering the
wavelength range $\lambda \sim 4900 - 7450$ \AA, with a mean dispersion in the
wavelength region of interest of $\approx 1.255$ \AA\ pixel$^{-1}$. 
Two exposures of 1200 s were obtained for SDSS 0338+0021, and four exposures
of 1200 s for RD 657, with the object dithered on the slit between the
exposures.  The slit PA was always close to the parallactic angle.
Arc lamps were used for wavelength calibration, with the flexure shifts
corrected using the measurements of selected night sky lines, and exposures
of standard star Hiltner 600 were used for the flux calibration.

We rebinned both grating data sets to a common sampling grid of 2 \AA\
pixel$^{-1}$, smoothed with Gaussians with $\sigma = 2$ \AA\ (lower than the
optimal smoothing filter for these data, thus resulting in no loss of
information).  Fig. 3 shows the combined spectra of the two objects.
The redshifts determined from the emission lines (taking into account the
absorption of the blue side of Ly$\alpha$) are $z_e = 5.02 \pm 0.02$ for
SDSS 0338+0021, and $z_e = 4.96 \pm 0.03$ for RD 657, with the large
uncertainties due to the intrinsic difficulty of centering of broad
emission lines.  We also note the presence of a Lyman limit system at
$z_{LLS} = 4.99$ in the spectrum of the brighter QSO.

While absolute values of the redshifts cannot be measured very precisely,
we used a simple cross-correlation to evaluate the redshift difference
between the two objects.  We excluded the portions of the spectra blueward
of the centroids of the Ly$\alpha$ lines, since the differences in the IGM
absorption between the two lines of sight could significantly affect the
results.  We obtain $\Delta z = 0.063 \pm 0.008$, which corresponds to the
restframe velocity difference of $\Delta v = 3150 \pm 400$ km s$^{-1}$
(these are the formal errors; possible systematic errors due to the mismatch
of the spectra are hard to estimate precisely, but could be of the same order).
Looking for correlated absorption systems in the two spectra would require
data with a higher S/N and a higher resolution.

We obtained preliminary spectra of a number of other, fainter, color-selected
candidates in the field.  While the results are still inconclusive, none of
them are luminous AGN.

\section{Discussion and Conclusions}

In what follows, for the sake of consistency with the previous work we will use
the ``standard quasar cosmology'' with $H_0 = 50$ km s$^{-1}$ Mpc$^{-1}$, 
$\Omega_0 = 1$, and $\Lambda_0 = 0$.  

In this cosmology, the observed angular separation of 196 arcsec corresponds to
a projected spatial separation of $1.126$ proper Mpc, or $6.744$ comoving Mpc
at $\langle z \rangle = 4.99$. 
If we assume that the observed $\Delta z = 0.063 \pm 0.008$ is due entirely to
cosmological expansion, the radial separation is $4.27 \pm 0.56$ proper Mpc
($25.6 \pm 3.6$ comoving). 
Quadratic sum of the two suggests a spatial separation of the two quasars of
$4.4 \pm 0.6$ proper Mpc ($26.5 \pm 3.6$ comoving). 
However, this interpretation of the observed redshift difference as being
due only to the Hubble expansion is uncertain.  If we did not have any radial
separation information, we could estimate the spatial separation from the
projected separation alone: statistically, for a pair randomly oriented in
space, the ratio of the two is $\pi/2$, so that the most probable spatial
separation corresponding to our projected separation is $1.77$ proper Mpc
($10.6$ comoving).  In the discussion below (which is similar to that by
\cite{sch00}) we will consider both of these possibilities, i.e., comoving 
separations of $26.5$ Mpc (the cosmological expansion model)
and $10.6$ Mpc (the deprojection model). 

Individual quasars represent rare events in the general population of galaxies
at any redshift; how likely is to have two of them so close together?
In order to estimate the probability of finding such a close quasar pair by
chance at this redshift, we use the evolving QSO luminosity function (QLF) by
\cite{fan01}.  For SDSS 0338+0021, we adopt the absolute blue magnitude 
$M_B = -26.56$ from \cite{fan99}.  Using the mean $i$ and $z$ magnitude
difference of $\approx 1.35$ mag, we estimate $M_B = -25.2$ for RD 657. 
The \cite{fan01} QLF gives a number density of $4.35 \times 10^{-8}$ Mpc$^{-3}$ 
for quasars with $M_B \leq -25.2$ at this redshift. 
The comoving volumes enclosed by spheres with radii equivalent to the physical
separations of the quasars of 10.6 and 25.6 Mpc are $4.97 \times 10^3$ Mpc$^3$
and $7.01 \times 10^4$ Mpc$^3$, respectively.

The first question we can ask is, what is the {\it a priori} probability of
finding such a close pair of QSOs at this redshift, regardless of the 
particular survey strategy?  Assuming a Poissonian distribution of quasars,
the probabilities of finding two QSOs at these luminosities in these volumes
are $2.3 \times 10^{-8}$ and $4.6 \times 10^{-6}$, respectively.  A similar
reasoning was applied to the two serendipitously discovered quasar pairs at
$z > 4$ (\cite{ssg94a}, \cite{sch00}). 

However, the volume in which we found this pair was not selected at random:
it is centered on a previously known QSO.  We can thus ask an alternative
question, which is specific for our experiment, namely, what is the probability
that another QSO is found at random in this particular volume?  (We note that
the same answer would apply whether or not there is an already known QSO in
its center.)  The probabilities then become $2.2 \times 10^{-4}$ and
$3.0 \times 10^{-3}$, respectively.  Thus, it is still unlikely that this
pair represents a chance occurence, suggesting that there is some physical
clustering present.

We note that as of early 2003, we observed a total of 14 fields of QSOs at
$z \gsim 4.8$ covering the FOV $\sim 25$ arcmin diameter ($\sim 52$ comoving
Mpc in the cosmology used here) to a comparable depth.  Our spectroscopic 
follow-up is still incomplete, and thus it would be premature to include this
additional volume in the present computation, but to date no other cases of
comparably bright QSO pairs have been found, suggesting that this system must
be a relatively rare event.  A full analysis and estimates of the QSO clustering
and bias will be presented in a future paper, once the survey is complete.

We also note that in a deeper Keck survey of $\sim 20$ QSOs at $z \sim 4 - 4.7$,
but covering a smaller FOV, $\sim 6 \times 8$ arcmin$^2$ ($\sim 12 \times 16$
comoving Mpc$^2$) we found at least 2 AGN companions to the known, bright QSOs,
with sub-Mpc separations (not gravitational lenses), as well as a number of
clustered faint galaxy companions (\cite{djorg99}, \cite{dj+99}, and in prep.).

The clustering strength cannot be meaningfully measured from a single
pair of objects in a survey of as yet poorly defined coverage.  With this
caveat in mind, the small probability of a random occurrence of such a pair
implies an effective $r_0$ which could be considerably greater than the
observed pair separation, i.e., $r_0 >> 10$ comoving Mpc.  
At low redshifts, there is some spread or results between
different groups (see, e.g., \cite{bm93}, \cite{cs96}, or \cite{sab01}),
but most authors find that the observed clustering length of quasars is
comparable to that of galaxy groups, $r_0 \sim 10 - 20$ Mpc (\cite{bc91}; see
\cite{hs90} for a review and references).  A standard parameterization of the
evolution of clustering in comoving coordinates is given by the formula 
$$ \xi (r,z) ~= ~\left( {{r}\over{r_0}} \right) ^{-\gamma}
                ~(1+z)^{-(3 -\gamma +\epsilon)} $$
\noindent
where $\gamma \approx 1.8$ and $\epsilon$ is the evolution parameter.  
For our chosen cosmology, the expected value for the CDM cosmogony is 
$\epsilon \approx 0.8$, and this is consistent with observations of the
evolution of galaxy clustering at $z < 1$ (see, e.g., \cite{carl00}).  Thus,
one expects a strong $decrease$ in the clustering strength at higher redshifts,
and in any model gravitational clustering is always expected to grow in time. 
How do we then explain the apparent $increase$ in the strength of quasar
clustering at high redshifts? 

The most natural explanation is that quasars represent highly biased peaks of
the density field, and that the bias itself evolves in time.
Ever since the first detections of QSO clustering (e.g., \cite{shav84},
\cite{sfbp87}, \cite{is88}, \cite{mf93}, etc.) it was considered possible
that QSOs represent biased tracers of the density field, but the evolution
of such bias was not clear.  \cite{lfac98} found a turn-up in the clustering
strength of quasars even at redshifts as low as $z \sim 2$, but this was not
confirmed in a much larger sample by \cite{croom01}.  \cite{out03} find no
evidence for an increase in the QSO clustering power spectrum amplitude
out to $z \sim 2.2$. 

The first hints of such an effect at high
redshifts were provided by the three few-Mpc quasar pairs at $z > 3$,
found in the statistically complete survey by \cite{ssg94b}, as first
pointed out by \cite{dts93}, and subsequently confirmed by more detailed
analysis (\cite{kun97}, \cite{step97}).  Intriguingly, the frequency of the
few-Mpc separation quasar pairs at lower redshifts is roughly what may be
expected from normal galaxy clustering (\cite{djorg91}; see also \cite{zs01}).
There is even a hint of a possible
superclustering of quasars at $z > 4$, on scales $\sim 100 ~h^{-1}$ comoving
Mpc (\cite{djorg98}), comparable to the scales of the first Doppler peak in
CMBR fluctuations.
Observations of large numbers of field galaxies at $z \sim 3 - 3.5$ also show a
relatively strong clustering, with $r_0 \sim 5 - 10$ Mpc, comparable to the
galaxy clustering
at $z \sim 0$ (\cite{stei98}, \cite{adel98}); this is also almost certainly a
manifestation of biasing. 

However, the bias should be even stronger at higher redshifts, and what is 
observed at $z \sim 3$ should be even more pronounced at $z \sim 5$:
the earliest massive galaxies, including quasar hosts, should be strongly
clustered.  An example may be the possible grouping of Ly$\alpha$ emitters
at $z \approx 4.86$ in the Subary Deep Field (\cite{ouch03}, \cite{shim03}).
Strong increase in biasing at high redshifts is also indicated in numerous
theoretical studies, e.g., by
\cite{bv94}, \cite{mclm97}, \cite{mclm98}, \cite{blan00}, \cite{mbmo00},
\cite{vss01}, \cite{bp01}, etc.  What these studies show is that a simple
$(r_0, \epsilon)$ parameterization of the clustering evolution is inadequate,
and that the evolution of the bias factor, $b$, plays a key role.  The
effective bias factor generally increases with the redshift and the object 
mass (e.g., especially for the more luminous Lyman-break galaxies or the quasar
hosts).  For example, \cite{croom02} find a marginally stronger clustering
for the brighter QSOs, which might be residing in more massive hosts, and
thus be more biased.

The chief uncertainty in our current understanding and interpretation of the
structure evolution at high redshift, as indicated by luminous objects we
can observe, is the evolution of bias.  Observations of the clustering of
quasars and galaxies around them at $z \sim 4 - 5$ and beyond can provide
valuable empirical constraints in this endeavor.  A better understanding of
the primordial clustering of luminous sources is also important for models
of the cosmic reionization (see \cite{dcsm01} and references therein), and thus
for the interpretation of CMBR fluctuations at high angular frequencies. 
The quasar pair described here may be indicative of the biased clustering at
$z \sim 5$, and more extensive and deeper surveys will provide additional
observational input for the models of galaxy and large-scale structure
formation.

\acknowledgments

We thank the anonymous referee for the insightful comments and suggestions.
We wish to acknowledge the expert assistance of the staff of the Palomar and
W.~M.~Keck Observatories.  The LRIS data were obtained in the course of a
collaborative project with F. Harrison and P. Mao.  SGD acknowledges partial
funding from the Ajax Foundation.  The work of DS was carried out at the
Jet Propulsion Laboratory, Caltech, under a contract with NASA.



\begin{figure*}[tbp]
\centerline{\psfig{file=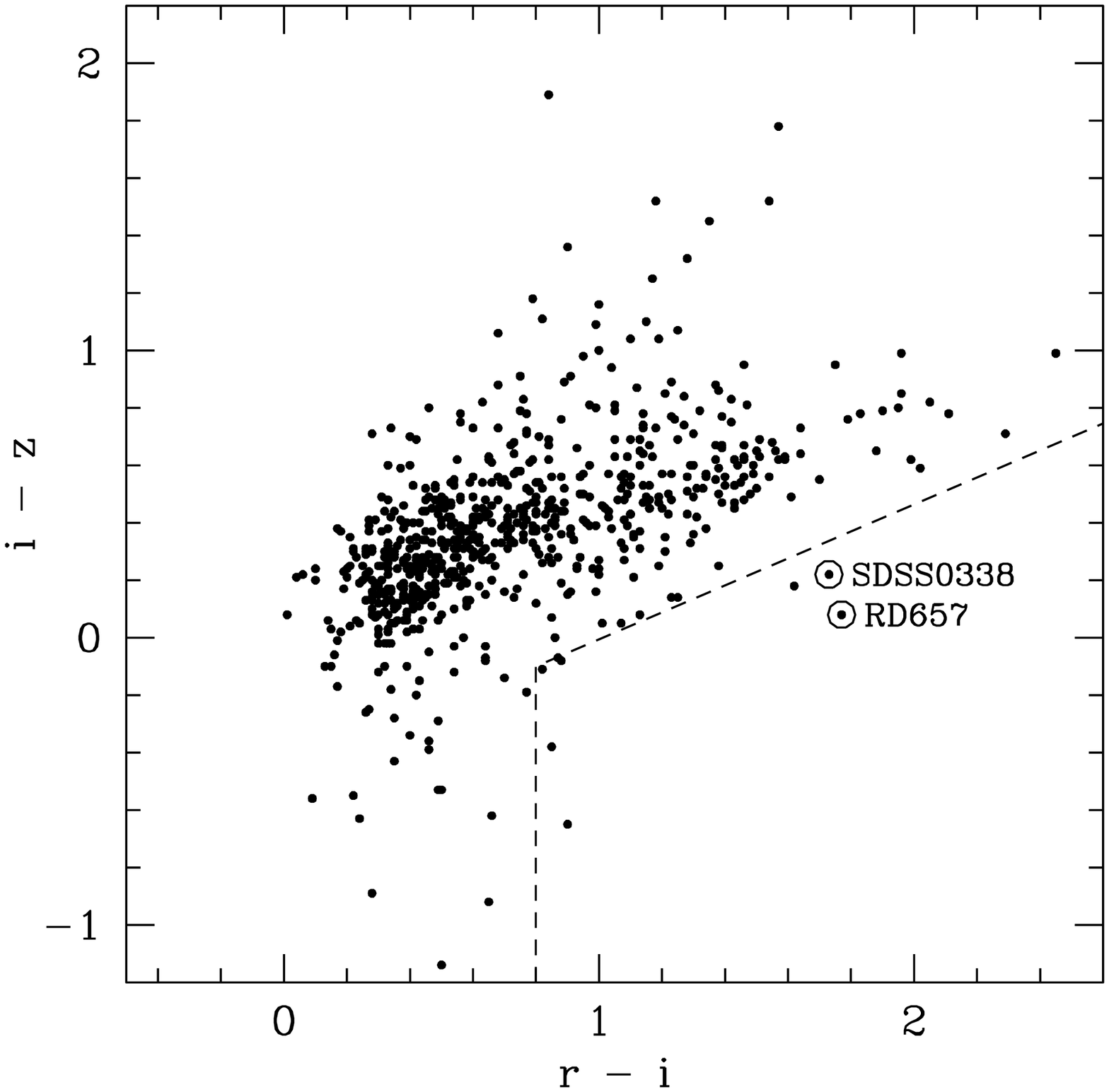,width=5.0in,angle=0}}
\caption[]{
\bf
A color-color diagram for all objects detected in the P200 imaging field, 
with the two quasars labeled.  The dashed line shows our color selection
boundary.  The two quasars are labeled.
}
\label{fig:fig1}
\end{figure*}


\begin{figure*}[tbp]
\centerline{\psfig{file=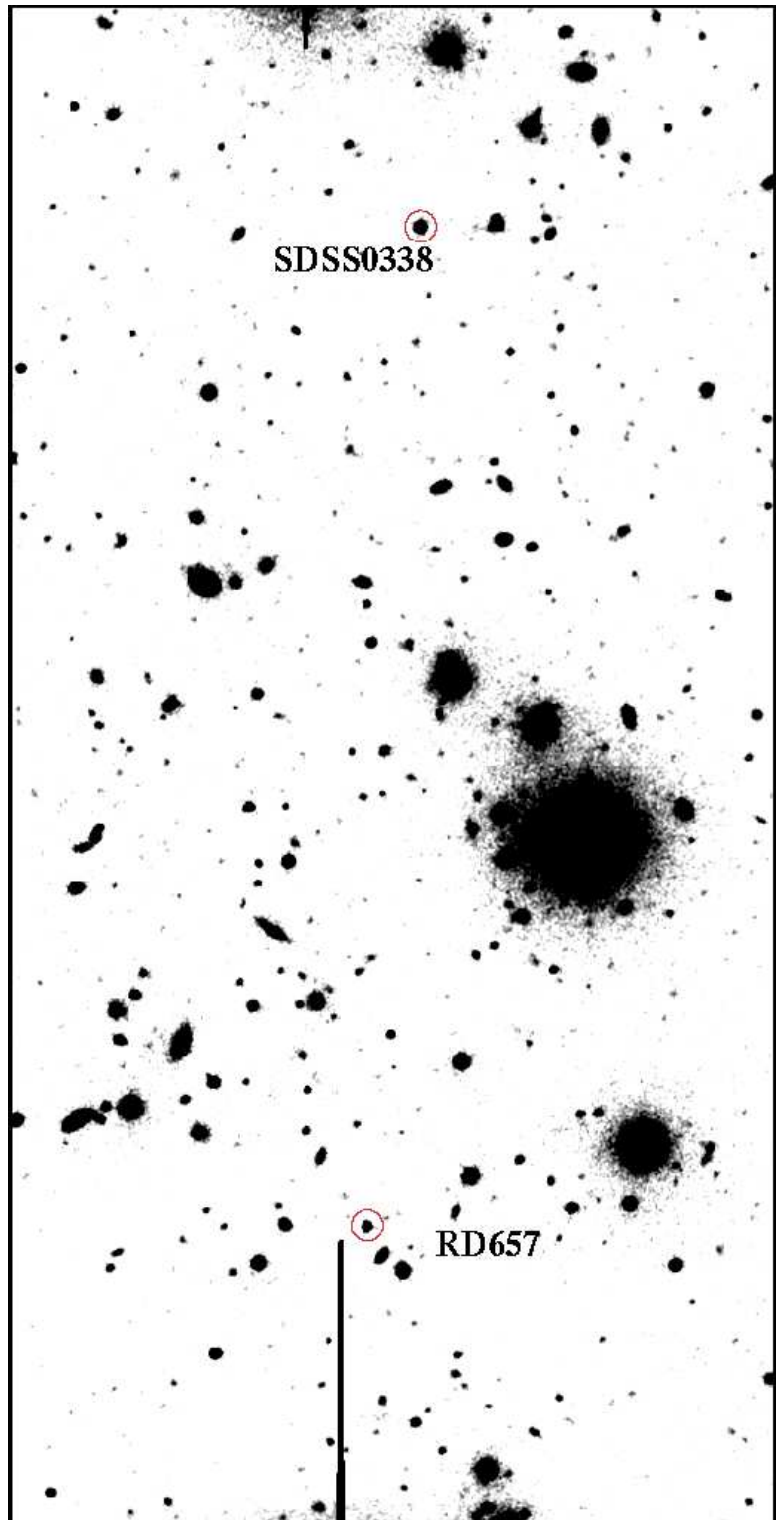,width=4.0in,angle=0}}
\caption[]{
\bf 
A finding chart for the field, from the P200 $i$-band images.  The field size
is 150 by 300 arcsec, with North up and East to the left.  The two quasars
are labeled.
}
\label{fig:fig2}
\end{figure*}


\begin{figure*}[tbp]
\centerline{\psfig{file=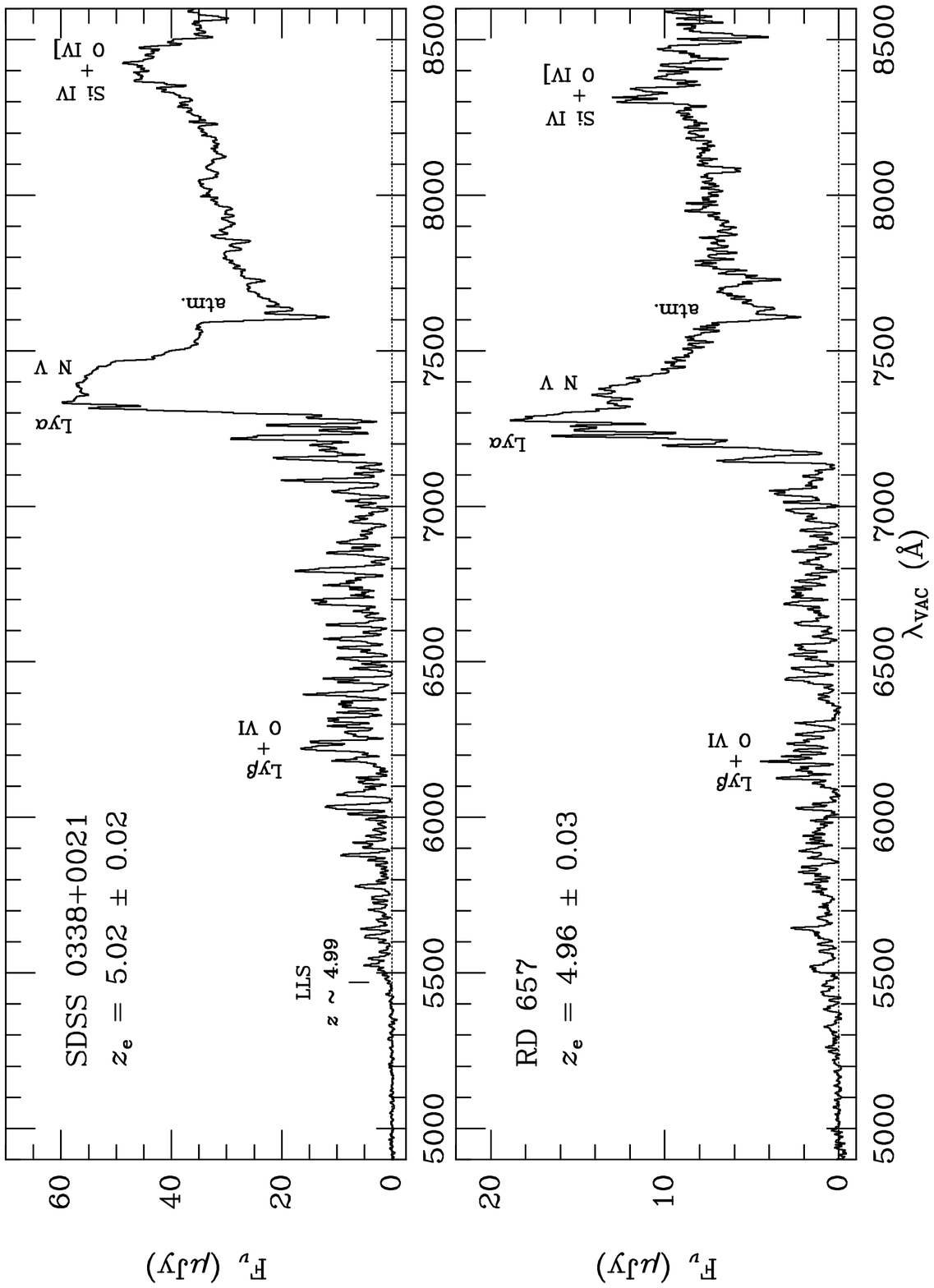,width=5.5in,angle=0}}
\caption[]{
\bf
Combined Keck/LRIS spectra of the two quasars, with the main features labeled.
}
\label{fig:fig3}
\end{figure*}


\begin{thebibliography}{}

\bibitem[Adelberger \etal 1998]{adel98} 
    Adelberger, K., Steidel, C., Giavalisco, M., Dickinson, M., Pettini, M., 
    \& Kellogg, M. 1998, ApJ, 505, 18

\bibitem[Bahcall \& Chokshi 1991]{bc91} 
    Bahcall, N., \& Chokshi, A. 1991, ApJ, 380, L9

\bibitem[Basilakos \& Plionis 2001]{bp01} 
    Basilakos, S., \& Plionis, M. 2001, ApJ, 550, 522

\bibitem[Blanton \etal 2000]{blan00} 
    Blanton, M., Cen, R., Ostriker, J., Strauss, M., \& Tegmark, M. 2000,
    ApJ, 531, 1

\bibitem[Boyle \& Mo 1993]{bm93} 
    Boyle, B., \& Mo, H.J. 1993, MNRAS, 260, 925
    
\bibitem[Brainerd \& Villumsen 1994]{bv94} 
    Brainerd, T., \& Villumsen, J. 1994, ApJ, 431, 477

\bibitem[Carlberg \etal 2000]{carl00} 
    Carlberg, R., Yee, H., Morris, S., Lin, H., Hall, P., Patton, D.,
    Sawicki, M., \& Shepherd, C. 2000, ApJ, 542, 57

\bibitem[Croom \& Shanks 1996]{cs96} 
    Croom, S., \& Shanks, T. 1996, MNRAS, 281, 893
    
\bibitem[Croom \etal 2001]{croom01} 
    Croom, S., Shanks, T., Boyle, B., Smith, R., Miller, L., Loaring, N.,
    \& Hoyle, F. 2001, MNRAS, 325, 483

\bibitem[Croom \etal 2002]{croom02} 
    Croom, S., Boyle, B., Loaring, N., Miller, L., Outram, P., 
    Shanks, T., \& Smith, R., 2001, MNRAS, 335, 459

\bibitem[Cole \& Kaiser 1989]{ck89} 
    Cole, S. \& Kaiser, N. 1989, MNRAS, 237, 1127

\bibitem[Djorgovski 1991]{djorg91} 
    Djorgovski, S.G. 1991, 
    in: The Space Distribution of Quasars, ed. D. Crampton,
    ASPCS, 21, 349

\bibitem[Djorgovski \etal 1993]{dts93} 
    Djorgovski, S., Thompson, D., \& Smith, J. 1993, 
    in: First Light in the Universe, eds. B. Rocca-Volmerange \etal,
    p.67, Gif sur Yvette: Eds. Fronti\`eres

\bibitem[Djorgovski 1998]{djorg98} 
    Djorgovski, S.G. 1998, 
    in: Fundamental Parameters in Cosmology, eds. Y. Giraud-Heraud \etal, 
    p.313, Gif sur Yvette: Eds. Fronti\`eres

\bibitem[Djorgovski \etal 1999]{dj+99} 
    Djorgovski, S.G., Odewahn, S.C., Gal, R.R., Brunner, R., \& de Carvalho, R.
    1999, in: Photometric Redshifts and the Detection of High Redshift Galaxies,
    eds. R. Weymann \etal, ASPCS, 191, 179

\bibitem[Djorgovski 1999]{djorg99}
    Djorgovski, S.G. 1999, 
    in: The Hy-Redshift Universe: Galaxy Formation and Evolution at High 
    Redshift, eds. A. Bunker \& W. van Breugel, ASPCS, 193, 397

\bibitem[Djorgovski \etal 2001]{dcsm01} 
    Djorgovski, S.G., Castro, S., Stern, D., \& Mahabal, A. 2001, ApJ, 560, L5

\bibitem[Efstathiou \& Rees 1988]{er98} 
    Efstathiou, G., \& Rees, M. 1988, MNRAS, 230, P5

\bibitem[Fan \etal 1999]{fan99}
    Fan, X., \etal (the SDSS Collaboration) 1999, AJ, 118, 1

\bibitem[Fan \etal 2001]{fan01}
    Fan, X., \etal (the SDSS Collaboration) 2001, AJ, 121, 54

\bibitem[Ferrarese \& Merritt 2000]{fm00} 
    Ferrarese, L., \& Merritt, D. 2000, ApJ, 539, L9

\bibitem[Franceschini \etal 1999]{fhmm99} 
    Franceschini, A., Hasinger, G., Miyaji, T., \& Malquori, D. 
    1999, MNRAS, 310, L5

\bibitem[Gebhardt \etal 2000]{geb00} 
    Gebhardt \etal 2000, ApJ, 539, L13

\bibitem[Granato \etal 2001]{gran01} 
    Granato, G., Silva, L., Monaco, P., Salucci, P., De Zotti, G., 
    \& Danese, L. 2001, MNRAS, 324, 757

\bibitem[Hamann \& Ferland 1999]{hf99} 
    Hamann, F., \& Ferland, G. 1999, ARAA, 37, 487

\bibitem[Hartwick \& Schade 1990]{hs90} 
    Hartwick, F.D.A., \& Schade, D. 1990, ARAA, 28, 437

\bibitem[Hoyle \etal 2002]{hoy02} 
    Hoyle, F., Outram, P., Shanks, T., Croom, S., Boyle, B., Loaring, N.,
    Miller, L., \& Smith, R. 2002, MNRAS, 329, 336

\bibitem[Iovino \& Shaver 1988]{is88} 
    Iovino, A., \& Shaver, P. 1988, ApJ, 330, L13

\bibitem[Kaiser 1984]{kai84} 
    Kaiser, N. 1984, ApJ, 284, L9

\bibitem[Kauffmann \& Haehnelt 2000]{kh00} 
    Kauffmann, G. \& Haehnelt, M. 2000, MNRAS, 311, 576

\bibitem[Kells \etal 1998]{kel98} 
    Kells, W., Dressler, A., Sivaramakrishnan, A., Carr, D., Koch, E.,
    Epps, H., Hilyard, D., \& Pardeilhan, G. 1998, PASP, 110, 1487

\bibitem[Kundic 1997]{kun97} 
    Kundic, T. 1997, ApJ, 482, 631

\bibitem[La Franca \etal 1998]{lfac98} 
    La Franca, F., Andreani, P., \& Cristiani, S. 1998, ApJ, 497, 529

\bibitem[Magliocchetti \etal 2000]{mbmo00} 
    Magliocchetti, M., Bagla, J., Maddox, S., \& Lahav, O. 2000, MNRAS, 314, 546
    MNRAS, 286, 115

\bibitem[Magorrian \etal 1998]{mag98} 
    Magorrian, J. \etal 1998, AJ, 115, 2285

\bibitem[Matarrese \etal 1997]{mclm97} 
    Matarrese, S., Coles, P., Lucchin, F., \& Moscardini, L. 1997, 
    MNRAS, 286, 115

\bibitem[Mo \& Fang 1993]{mf93} 
    Mo, H.J., \& Fang, L.Z. 1993, ApJ, 410, 493
    
\bibitem[Monaco \etal 1999]{msd00} 
    Monaco, P., Salucci, P., \& Danese, L. 2000, MNRAS, 311, 279

\bibitem[Moscardini \etal 1998]{mclm98} 
    Moscardini, L., Coles, P., Lucchin, F., \& Matarrese, S. 1998, 
    MNRAS, 299, 95

\bibitem[Nusser \& Silk 1993]{ns93} 
    Nusser, A., \& Silk, J. 1993, ApJ, 411, L1

\bibitem[Oke \etal 1995]{occ+95}
    Oke, J.~B. \etal 1995, PASP, 107, 375

\bibitem[Ouchi \etal 2003]{ouch03} 
    Ouchi, M. \etal 2003, ApJ, 582, 60

\bibitem[Outram \etal 2003]{out03} 
    Outram, P., Hoyle, F., Shanks, T., Croom, S., Boyle, B., Miller, L., 
    Smith, R., \& Myers, A. 2003, MNRAS, in press

\bibitem[Sabbey \etal 2001]{sab01} 
    Sabbey, C., Oemler, A., Coppi, P., Bongiovanni, A., Bruzual, G., Garcia, C.,
    Musser, J.,  Rengstorf, A., \& Snyder, J. 2001, ApJ, 548, 585
    
\bibitem[Schneider \etal 1994a]{ssg94a} 
    Schneider, D., Schmidt, M.,  \& Gunn, J. 1994, AJ 107, 880

\bibitem[Schneider \etal 1994b]{ssg94b} 
    Schneider, D., Schmidt, M.,  \& Gunn, J. 1994, AJ 107, 1245

\bibitem[Schneider \etal 2000]{sch00} 
    Schneider, D., \etal (the SDSS Collaboration) 2000, AJ, 120, 2183

\bibitem[Shanks \etal 1987]{sfbp87} 
    Shanks, T., Fong, R., Boyle, B., \& Peterson, B. 1987, MNRAS, 227, 739
    
\bibitem[Shaver 1984]{shav84} 
    Shaver, P. 1984, A\&A, 136, L9

\bibitem[Shimasaku \etal 2003]{shim03} 
    Shimasaku, K. \etal 2003, ApJ, 586, L111

\bibitem[Silk \& Rees 1998]{sr98} 
    Silk, J., \& Rees, M. 1998, A\&A, 331, L1

\bibitem[Songaila \etal 1999]{shcm99}
    Songaila, A., Hu, E., Cowie, L., \& McMahon, R. 1999, ApJ, 525, L5

\bibitem[Steidel \etal 1998]{stei98} 
    Steidel, C., Adelberger, K., Dickinson, M., Pettini, M., \& Kellogg, M. 
    1998, ApJ, 492, 428 


\bibitem[Stephens \etal 1997]{step97} 
    Stephens, A., Schneider, D., Schmidt, M., Gunn, J., \& Weinberg, D. 1997, 
    AJ, 114, 41


\bibitem[Turner 1991]{tur91} 
    Turner, E. 1991, AJ, 101, 5

\bibitem[Valageas \etal 2001]{vss01}
    Valageas, P., Silk, J., \& Schaefer, R. 2001, A\&A, 366, 363

\bibitem[Zhdanov \& Surdej 2001]{zs01}
    Zhdanov, V., \& Surdej, J. 2001, A\&A, 372, 1

\end{thebibliography}
\end{document}